# Relationship between spin-glass three-dimensional (3D) Ising model and traveling salesman problems

Zhidong Zhang

Shenyang National Laboratory for Materials Science, Institute of Metal Research, Chinese Academy of Sciences, 72 Wenhua Road, Shenyang, 110016, P.R. China

**Abstract**

In this work, the relationship between a spin-glass three-dimensional (3D) Ising model with the lattice size $N = mnl$ and the traveling salesman problem (TSP) in a 3D lattice is studied. In particular, the mathematical structures of the two systems are investigated in details. In both the hard problems, the nontrivial topological structures, the non-planarity graphs, the nonlocali**ties** and/or the long-range spin entanglements exist, while randomness presents, which make the computation very complicated. It is found that an absolute minimum core (AMC) model $M_{AMC}^{3D}$ exists not only in the spin-glass 3D Ising model but also in the 3D TSP for determining the lower bound of their computational complexities, which can be mapped each other. That is, $M_{AMC,SGI}^{3D} \Leftrightarrow M_{AMC,TSP}^{3D}$ and for computational complexities $C(M_{AMC,SGI}^{3D}) = C(M_{AMC,TSP}^{3D})$. It is verified that the spin-glass AMC model $M_{AMC,SGI}^{3D}$ equals to the difference between a two-level ($l = 2$) grid spin-glass 3D Ising model and a spin-glass 2D Ising model, namely. $M_{AMC,SGI}^{3D} = M_{l=2,SGI}^{3D} - M_{SGI}^{2D}$, which is NP-complete. Furthermore, according to the mapping between the spin-glass 3D Ising model and the TSP, it is proven that the AMC model $M_{AMC,TSP}^{3D}$ for the TSP identifies to the difference between a two-level ($l = 2$) grid TSP model and a 2D TSP model. Namely,

$M_{AMC,TSP}^{3D} = M_{I=2,TSP}^{3D} - M_{TSP}^{2D}$, which is NP-complete also. The AMC models in both models are proven to be at the border between the NP-complete problems and the NP-intermediate problems. Because the lower bound of the computational complexity of the spin-glass 3D Ising model $C_L(M_{SGI}^{3D})$ is the computational complexity by brute force search of the AMC model $C^U(M_{AMC,SGI}^{3D})$, the lower bound of the computational complexity of the TSP in a 3D lattice, $C_L(M_{TSP}^{3D})$, is the computational complexity by brute force search of the AMC model for the TSP, $C^U(M_{AMC,TSP}^{3D})$. Namely, $C_L(M_{TSP}^{3D}) = C_L(M_{SGI}^{3D}) \geq C^U(M_{AMC,SGI}^{3D}) = C^U(M_{AMC,TSP}^{3D})$. All of them are in subexponential and superpolynomial. The present work provides some implications on numerical algorithms for the NP-complete problems, for instance, one cannot develop a polynomial algorithm, but may develop a subexponential algorithm for the 3D spin-glass problem or TSP.

The corresponding author: Z.D. Zhang, Tel: 86-24-23971859, Fax: 86-24-23891320, e-mail address: zdzhang@imr.ac.cn

1. Introduction

There has been a great evidence that many hard problems in different fields, such as, mathematics, physics, chemistry, biology, computer sciences, etc., are connected in depth levels. To explicitly solve a difficult problem in physics, one may need to understand well the mathematical structures with algebraic, topological and geometric aspects of this physical system. On the other hand, to solve a hard problem in mathematics and/or computer sciences, one may need to utilize physical thoughts to get a significance to guide the process of proving theorems. Furthermore, solving successfully a problem in either physics or mathematics can be extended to be applicable in other fields (for instance, chemistry, biology, computer sciences, and even society sciences).

In computational complexity theory, non-deterministic polynomial time is abbreviated as NP. It is defined as the set of decision problems that can be solved in polynomial time on a non-deterministic Turing machine. NP-complete problems (Cook-Levin theorem [11,28]) are NP-hard problems cataloged to the set of all decision problems in NP, whose solutions can be verified in polynomial time. If every other problem in NP can be reduced into p in polynomial time, then a problem p in NP is NP-complete. The traveling salesman problem (TSP), sometime called Chinese postman problem [4,26], is a well-known hard problem in computer sciences, being a central problem in combinatorial optimization [2,10,12,14,15,20,32,37,43]. The TSP is defined as follows [5,33,36,45,50]. N cities are located at points $R_i$, i = 1, 2 ..... N, in a D-dimensional space. A traveling salesman has to visit all of the cites and return to the starting point at the end of the tour. Taking into account the two traversals (in opposite directions) of each tour and the arbitrariness of the starting city, there are (N-1)!/2 distinct tours. The TSP is asking to find the shortest tour(s) (the optimal one) among them. It should be noticed that the computational complexity of the TSP relies on its dimensionality. If the tours are limited to a trivial graph in a plane, the optimal problem for the TSP will be a P-problem. If the tours allow a nontrivial graph with crossings, the optimal problem for the TSP will be a NP-complete problem.

An efficient algorithm for solving the TSP in its worst-case instances will immediately lead to other algorithms for solving efficiently thousands of different hard combinatorial problems. Among thousands of them, we just mention several well-known NP-complete problems as follows [6,8,9,16,17,18,25,35,44,49]: Boolean satisfiability (K-SAT) problem, Hamiltonian path problem, Knapsack problem, subset sum problem, vertex cover problem, independent set problem, graph coloring problem, protein folding problem, etc. Up to date, all known algorithms for NP-complete problems require time that is superpolynomial in the input size.

The spin-glass three-dimensional (3D) Ising model is related closely with graph theory, combinatorial optimization and statistical learning networks [1,3,4,7,13,17,19,34,47]. The spin-glass 3D Ising model is one of the NP-complete problems. In a spin glass, all the spins are frozen in a disorder ground state, aligning randomly to different directions. The spin-glass 3D Ising model describes the spin-glass system by considering randomly distributed interactions between spins (with a mixture of roughly equal numbers of ferromagnetic and antiferromagnetic bonds). One may view the spin-glass state as an ordered state with disorder orientations of spins, frozen from a disorder paramagnetic state with decreasing temperature, at a phase transition associated to the onset of the spontaneous replica symmetry breaking. Moreover, some bonds may satisfy the frustration requirements on elementary plaquettes in the spin-glass model, which would cause the worst case for spin configurations and for computational complexity [1,3,4,7,13,17,19,34,47].

Although the NP-complete problem is one of the most important problems in mathematics and computer sciences, it involves thousands of problems in different fields, such as physics, chemistry, biology and so on. It is accessible for mathematicians and computer scientists to eager to learn the main ideas and tools of statistical physics when applied to random combinatorics. Actually, there have been many examples that progress in one discipline can benefit the others. Thus, any advances in these variation fields for anyone of these problems may shed a light on solving the NP-complete problem. The present author has been working on the 3D Ising models for tens of years and figures out the important feature of the mathematical structures [48,52-59], which are quite helpful for understanding the character of the spin-glass 3D Ising model, and thus also for the NP-complete problem. It is worth mentioning recent Monte Carlo simulations [29,30], in which the critical exponents of the 3D Ising model obtained by taking into account the effect of long-range interactions of spin chains (namely, the nontrivial topological contribution) agree well with my exact solution. In a previous work [56], the computational complexity of a spin-glass 3D Ising model was studied, and its lower bound was determined to be in subexponential time and superpolynomial time. In another work [60], the mapping between a spin-glass 3D Ising model and Boolean satisfiability problem was established by duality relations. In a recent work [61], the lower bound of computational complexity of Knapsack problems was determined. The aim of this work is to investigate the mapping between the spin-glass 3D Ising model and the TSP, and to investigate in details their mathematical structures, in particular, by use of schematic illustrations. At first, following our previous work [56,60,61], we prove that an absolute minimum core (AMC) model exists in the spin-glass 3D Ising model for determining the lower bound of its computational complexity. Utilizing the relationship between the TSP and the spin-glass 3D Ising model, we then prove that an AMC model does exist also in the TSP. We prove that the AMC models in both the models are NP-complete, and located at the border between the NP-complete problems and the NP-intermediate (NPI) problems. Furthermore, we prove that the lower bound of the computational complexity of the TSP in a 3D lattice $C_L(M_{TSP}^{3D})$ is the computational complexity by brute force search of the AMC model for the TSP, $C^U(M_{AMC,TSP}^{3D})$. Namely, $C_L(M_{TSP}^{3D}) \geq C^U(M_{AMC,TSP}^{3D})$. These computational complexities are in subexponential and superpolynomial.

## 2. Mathematical structures of the spin-glass 3D Ising model

**Definition 1.** *Let $M_A^D$ be a physical model where the upper script fixes the dimension, and the lower indices indicate the character of the model.*

**Definition 2.** *Let $C(M_A^D)$ be the computational complexity of the model $M_A^D$.*

**Definition 3.** *Let $C^U(M_A^D)$ be the upper bound of the computational complexity of $M_A^D$. The upper bound for a model equals to the computational complexity as computed by brute force search.*

**Definition 4.** *Let $C_L(M_A^D)$ be the lower bound of the computational complexity of $M_A^D$.*

**Definition 5.** *The absolute minimum core (AMC) model of the spin-glass 3D Ising model, $M_{AMC,SGI}^{3D}$, is defined as a spin-glass 2D Ising model interacting with its nearest neighboring plane.*

According to the procedure in [56,60,61], an AMC model exists in the spin-glass 3D Ising model. Here, the mathematical structures of the spin-glass 3D Ising model and the AMC model are investigated in details, particularly, illustrated by schematic figures.

The Hamiltonian of a spin-glass 3D Ising model, $M_{SGI}^{3D}$, is written as [13,21,23,38, 52-58]:

$$H = -\sum_{<i,j>} J_{ij} S_i S_j$$

(1)

where Ising spins with $S = 1/2$ are arranged on a 3D lattice with the lattice size $N = mnl$. The numbers (*m*, *n*, *l*) denote lattice points along three crystallographic directions. We consider the Edwards–Anderson model with only the nearest neighboring interactions $J_{ij}$ with different signs, which are randomly distributed and can be set to be different. We use $\tilde{J}$, $\tilde{J}'$ and $\tilde{J}''$ to represent the randomly distributed interactions along three crystallographic directions, respectively. For solving the exact solution of the free energy and thermodynamic properties and finding the ground state of the spin-glass 3D Ising model, one must take the thermodynamic limit $m \to \infty$, $n \to \infty$, $l$

→ ∞, $N \to \infty$. Figure 1 illustrates schematically an example of a spin-glass 3D Ising lattice with $m = n = l = 8$. Notice that frustrations may exist among spins in some plaquettes in the spin-glass system (for simplicity, frustrations are not illustrated in the figure).

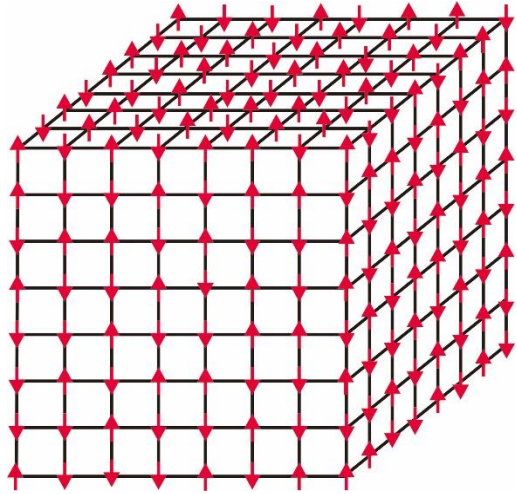

Figure 1. Schematic illustration of a spin-glass 3D Ising model, $M_{SGI}^{3D}$, in which spins (red arrows) located at every lattice point of a 3D lattice (with the lattice size $N = mnl$, here $m = n = l = 8$ as an example, black lines represent the lattice) align along with randomly distributed directions, which are caused by randomly distributed interactions between spins. For simplicity, frustrations which may exist among spins in some plaquettes are not illustrated.

**Theorem 1.** *The spin-glass 3D Ising model, $M_{SGI}^{3D}$, is NP-complete [1,3,22].*

**Proof.** As usual, the probability of finding the spin-glass 3D Ising lattice in a given configuration, and a fixed replica, at the temperature *T*, is proportional to $\exp\left\{-\frac{E_c}{k_B T}\right\}$, where $E_c$ is the total energy of the configuration and $k_B$ is the Boltzmann constant. The thermodynamic properties for the spin-glass 3D Ising model can be found from the partition function Z, after mediating $\overline{lnZ}$ over disorder. The partition function $\bar{Z}_\alpha$ for the spin-glass 3D Ising lattice can be expressed in a fixed replica α (= 1,2,…R) as [13,21,23,31,38,52]:

$$\bar{Z}_\alpha = \sum_{all\ configurations} e^{n_c \widetilde{K} + n'_c \widetilde{K}' + n''_c \widetilde{K}''}$$

(2)

Here we use $\bar{Z}_\alpha$ to represent the partition function in a fixed replica. The partition function Z can be calculated from the product of the partition function of all fixed replicas. For the lower bound of the computational complexity, it is enough to focus on $\bar{Z}_\alpha$, due to a fact that the computational complexity for computing the partition function Z is much more complicated than that of $\bar{Z}_\alpha$. $n_c$, $n'_c$ and $n''_c$ are integers depending on the configuration of the spin lattice. The variables $\widetilde{K} \equiv \tilde{J}/(k_B T)$, $\widetilde{K}' \equiv \tilde{J}'/(k_B T)$ and $\widetilde{K}'' \equiv \tilde{J}''/(k_B T)$, are introduced instead of $\tilde{J}$, $\tilde{J}'$ and $\tilde{J}''$ for the randomly distributed interactions. The partition function $\bar{Z}_\alpha$ of the spin-glass 3D Ising lattice in a fixed replica may be written in forms of three transfer matrices in forms of direct products of matrices [21,23,31,38,42,48,52-58]. The following generators of Clifford algebra of the 3D Ising model are introduced:

$$\Gamma_{2k-1} = C \otimes C \otimes \ldots \ldots \otimes C \otimes s' \otimes 1 \ldots \otimes 1 \quad \text{(k-1 times } C\text{)} \tag{3}$$

$$\Gamma_{2k} = C \otimes C \otimes \ldots \ldots \otimes C \otimes (-is'') \otimes 1 \ldots \otimes 1 \quad \text{(k-1 times } C\text{)} \tag{4}$$

Following the Onsager-Kaufman-Zhang notation [21,23,38,48,52-58], we have: $s'' = \begin{bmatrix} 0 & -1 \\ 1 & 0 \end{bmatrix}$ (= $i\sigma_2$), $s' = \begin{bmatrix} 1 & 0 \\ 0 & -1 \end{bmatrix}$ (= $\sigma_3$), $C = \begin{bmatrix} 0 & 1 \\ 1 & 0 \end{bmatrix}$ (= $\sigma_1$), where $\sigma_j$ (j = 1,2,3) are Pauli matrices. The partition function $\bar{Z}_\alpha$ of the spin-glass 3D Ising model in a fixed replica can be expressed as follows [31,42,48,52-58]:

$$\bar{Z}_\alpha = (2sinh2\widetilde{K})^{\frac{mnl}{2}} \cdot \text{trace}(V_3 V_2 V_1) \tag{5}$$

$$V_3 = \prod_{j=1}^{mnl} exp\left\{-i\widetilde{K}''\Gamma_{2j}\left[\prod_{k=j+1}^{j+mn-1} i\Gamma_{2k-1}\Gamma_{2k}\right]\Gamma_{2j+2mn-1}\right\} = \prod_{j=1}^{mnl} exp\{i\widetilde{K}''s'_j s'_{j+mn}\}$$

(6)

$$V_2 = \prod_{j=1}^{mnl} exp\{-i\widetilde{K}'\Gamma_{2j}\Gamma_{2j+1}\} = \prod_{j=1}^{mnl} exp\{i\widetilde{K}'s'_j s'_{j+1}\}$$

(7)

$$V_1 = \prod_{j=1}^{mnl} exp\{i\widetilde{K}^* \cdot \Gamma_{2j-1}\Gamma_{2j}\} = \prod_{j=1}^{mnl} exp\{i\widetilde{K}^* \cdot C_j\}$$

(8)

Here $\widetilde{K}^*$ is defined by $e^{-2\widetilde{K}} \equiv tanh\widetilde{K}^*$ [21,23,31,38,42,48,52-58]. We define the matrices $C_j$ and $s'_j$ as follows:

$$C_j = I \otimes I \otimes \ldots\ldots \otimes I \otimes C \otimes I \otimes \ldots \otimes I$$

(9)

and

$$s'_j = I \otimes I \otimes \ldots\ldots \otimes I \otimes s' \otimes I \otimes \ldots \otimes I$$

(10)

It was proven that the spin-glass 3D Ising model, $M_{SGI}^{3D}$, is NP-complete [1,3,22], due to the existence of the non-planarity graphs. Indeed, the nonlinear terms of the internal factors as well as the terms of $s'_j s'_{j+mn}$ in the transfer matrix $V_3$ (see Eq. (6)) indicate the existence of the nontrivial topological structures, the non-planarity graphs, the nonlocalit**ies** and the long-range spin entanglements in the spin-glass 3D Ising model. All these characters together with randomness of interactions and spin alignments (and also frustrations) make the system to be NP-complete. □

A spin-glass 3D Ising lattice can be constructed by stacking $l$ layers of spin-glass 2D Ising lattices. This is the simplest way to construct the spin-glass 3D Ising model layer by layer, while keeping the characters and (thus the physical properties) of the spin-glass 3D (and also 2D) Ising model. Actually, other ways of constructions may cause more complicated procedures (referred to Theorem 2 in [56]). A natural equation is: If we reduced the layer numbers $l$, to what context the spin-glass 3D Ising lattice is still NP-complete? Namely, what is the limit in the parametric space of the spin-glass systems for the NP-complete problems? Or what is the border between the NP-complete problem and the P-problem? In the previous work [3], it was proven that the two-level ($l = 2$) grid spin-glass Ising model, $M^{3D}_{l=2,SGI}$, is NP-complete (see Theorem 2 below). In our previous work [56,60,61], we found that an AMC model exists in the spin-glass 3D Ising model, which is located at the border between the NP-complete problems and the NPI problems. In what the follows, we shall illustrate these results schematically.

Figure 2 schematically illustrates an example for a two-level ($l = 2$) grid spin-glass 3D Ising model, $M^{3D}_{l=2,SGI}$.

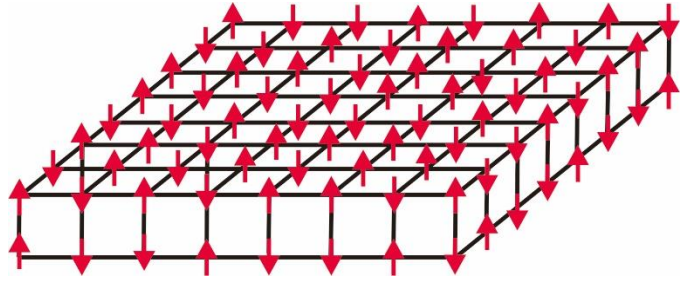

Figure 2. Schematic illustration of a two-level grid spin-glass 3D Ising model, $M^{3D}_{l=2,SGI}$, in which spins (red arrows) located at every lattice point of a two-level grid lattice (with the lattice size $N = mnl$, here $m = n = 8$ and $l = 2$ as an example, black lines represent the lattice) align along with randomly distributed directions, which are caused by randomly distributed interactions between spins. For simplicity, frustrations which may exist among spins in some plaquettes are not illustrated.

**Theorem 2.** *The two-level grid spin-glass Ising model, $M^{3D}_{l=2,SGI}$, is NP-complete [3].*

**Proof.** It has been proven that the spin-glass Ising model on a graph with a two-level grid [3,19,51] belongs to the class of NP-hard problems (schematically illustrated in Figure 2). □

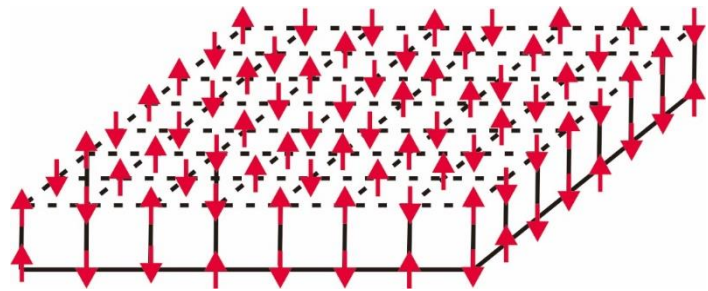

Figure 3. Schematic illustration of an AMC model for the spin-glass 3D Ising model, $M^{3D}_{AMC,SGI}$, in which spins (red arrows) located at every lattice point of a two-level grid lattice (with the lattice size $N = mnl$, here $m = n = 8$ and $l = 2$ as an example)) align along with randomly distributed directions, caused by randomly distributed

interactions between spins. Notice that we need two layers to represent the AMC model, in which the solid lines represent the bottom ($l = 1$) layer with the intralayer interactions and the interlayer interaction between the two layers, while the dashed lines show that there are no intralayer interactions on the top layer ($l = 2$).

**Theorem 3.** *An AMC model, $M_{AMC,SGI}^{3D}$, exists in the spin-glass 3D Ising model for determining the lower bound of its computational complexity. We have* $C_L(M_{SGI}^{3D}) \geq C^U(M_{AMC,SGI}^{3D})$.

**Proof.** In our previous work for solving exactly the ferromagnetic 3D Ising model [48,52-58], we have already illustrated that two contributions exist for physical properties of the system. One comes from the local spin alignments, while another comes from the nontrivial topological structures. The latter corresponds to the nonlinear internal factors in the transfer matrices $V_3$. As revealed in Eq. (6) above, the nontrivial topological structures exist also in the spin-glass 3D Ising model, which represent the long-range entanglements between spins in a plane (that is, terms of $s'_j s'_{j+mn}$). It was proven in [56,60,61] that an AMC model, $M_{AMC,SGI}^{3D}$, exists in the spin-glass 3D Ising model. According to Theorem 2 of [60], any algorithms, which use any approximations and/or break the long-range spin entanglements in the AMC model, $M_{AMC,SGI}^{3D}$, cannot result in the exact solution of the spin-glass 3D Ising model, $M_{SGI}^{3D}$. According to Theorem 3 of [60], we have $C_L(M_{SGI}^{3D}) \geq C^U(M_{AMC,SGI}^{3D})$. □

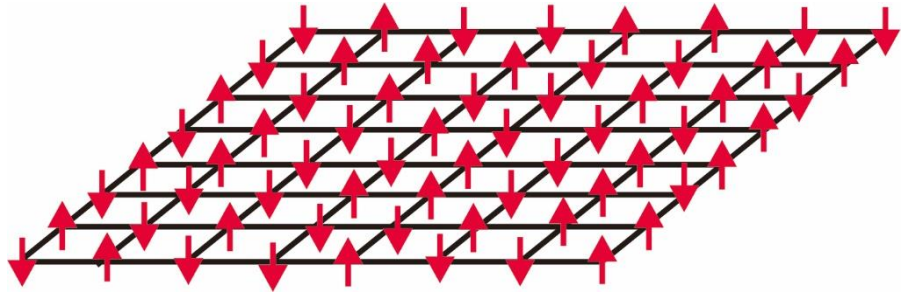

Figure 4. Schematic illustration of a spin-glass 2D Ising model, $M_{SGI}^{2D}$, in which spins (red arrows) located at every lattice point of a 2D lattice (with the lattice size $N = mnl$, here $m = n = 8$ and $l = 1$ as an example, black lines represent the lattice) align along with randomly distributed directions, which are caused by randomly distributed interactions between spins.

**Theorem 4.** *The spin-glass 2D Ising model, $M_{SGI}^{2D}$, is P-problem [3,4].*

**Proof.** It was proven in [3,4] that the spin-glass 2D Ising model, $M_{SGI}^{2D}$, is P-problem.

□

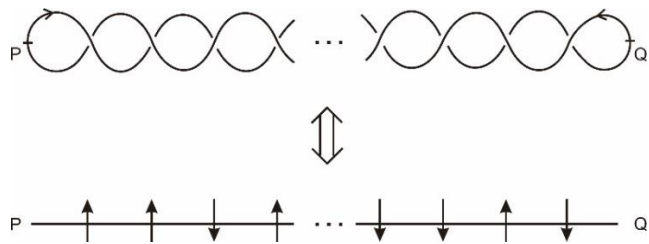

Figure 5. Mapping between a braid with randomly distributed crossings and a spin chain with randomly distributed alignments of spins.

Following the procedure in [58], the entanglement between spins in a plane (terms of $s'_j s'_{j+mn}$) can be illustrated by either a braid with many crossings or a spin chain with many spins. Figure 5 shows the mapping between a braid with randomly distributed crossings and a spin chain with randomly distributed alignments of spins.

For the ferromagnetic case, the crossings in a braid have the same cross and the spins are aligned in regularity (see Figure 4 in [58]). For the present problem of the spin-glass models, the crossings in a braid have the randomly distributed crosses and the spins in a spin chain are aligned in randomness (see Figure 5 above).

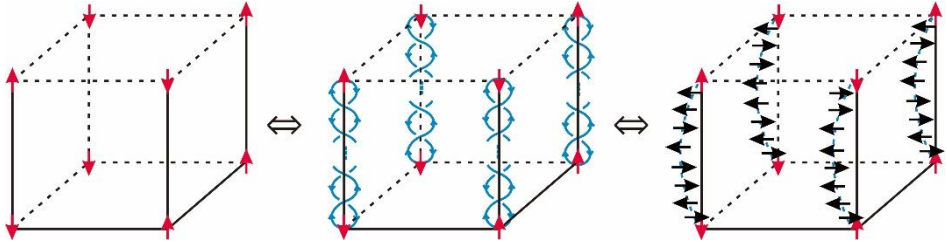

Figure 6. Schematic illustration of mapping in a unit cell of a cubic Ising lattice. The left figure shows an original illustration for the unit cell of the spin-glass 3D Ising model, $M_{SGI}^{3D}$, in which spins (red arrows) are located at every lattice site. In the middle figure, besides spins (red arrows) located at every lattice site, a braid (blue curves) is attached to connect a pair of the two nearest neighboring sites along the third dimension. In the right figure, besides spins (red arrows) located at every lattice site, spin chains with randomly aligned spin directions (black arrows) are attached to every pair of the two nearest neighboring lattice sites along the third dimension.

To simplify the illustration, we use a unit cell to represent a spin-glass 3D Ising model (the left figure in Figure 6), $M_{SGI}^{3D}$, in which spins (red arrows) are located at every lattice site. If we reproduced the unit cell along three crystallographic directions, while keeping the randomness of spin alignments, we would construct the spin-glass 3D Ising model, $M_{SGI}^{3D}$. If we reproduced the unit cell along only two crystallographic directions, while keeping the randomness of spin alignments, we would construct two-level grid spin-glass 3D Ising model, $M^{3D}_{l=2,SGI}$. In the middle figure of Figure 6, besides spins located at every lattice site, a braid is attached to connect a pair of the two nearest neighboring sites along the third dimension. The spins represent the linear terms in the transfer matrices, while the braids represent the nontrivial knots, namely, the nonlinear terms of Γ-matrices of the transfer matrices, of the 3D Ising lattice. In the right figure of Figure 6, besides spins located at every lattice site, spin chains with randomly aligned spin directions are attached to every pair of the two nearest neighboring lattice sites along the third dimension. The latter structure is derived from the former one by the mapping illustrated in Figure 5.

**Theorem 5.** $M^{3D}_{AMC,SGI} = M^{3D}_{l=2,SGI} - M^{2D}_{SGI}$, *which is NP-complete.*

**Proof.** To illustrate the AMC model, $M^{3D}_{AMC,SGI}$, we need two layers, since the transfer matrices for an AMC model involve the states of the spins in the two layers. This is the basic element of the spin-glass 3D Ising model, $M^{3D}_{SGI}$, which cannot be broken. Figure 3 schematically illustrates an AMC model in the spin-glass 3D Ising model, $M^{3D}_{AMC,SGI}$, in which randomly aligned spins are located at lattice points of the two nearest neighboring planes. Compared with a two-level grid spin-glass Ising model, $M^{3D}_{l=2,SGI}$, one finds that the AMC model does not take into account the intralayer interactions between the spins on the top layer. The dashed line on the top layer (l = 2) in Figure 3 shows that it is a neighboring plane for the layer with $l = 1$, to interact with along the third dimension, but without any intralayer interactions between spins on this layer ($l = 2$). Since the interactions between spins on one layer are described by a spin-glass 2D Ising model, $M^{2D}_{SGI}$, we have the identification $M^{3D}_{AMC,SGI} =$

$M_{l=2,SGI}^{3D} - M_{SGI}^{2D}$. According to Theorem 2, $M_{l=2,SGI}^{3D}$ is NP-complete and according to Theorem 4, $M_{SGI}^{2D}$ is P-problem. Therefore, $M_{AMC,SGI}^{3D}$ is NP-complete. □

It should be mentioned here that according to Theorem 1 in [56], the computational complexity of the core model of the spin-glass 3D Ising model is much higher than that of the AMC model. This is because in some replicas, frustration in the 3D case could appear on closed polygons, which are higher than a plaquette and cannot be included always in two neighboring planes (namely, more neighboring planes must be considered, if we consider all the possible frustrations in the 3D lattice).

**Theorem 6.** *The NPI problem exists in between $M_{AMC,SGI}^{3D}$ and $M_{SGI}^{2D}$.*

**Proof.** We have proven [56,60,61] that the lower bound of the computational complexity of the spin-glass 3D Ising model $C_L(M_{SGI}^{3D})$ is the computational complexity by brute force search of the AMC model $C^U(M_{AMC,SGI}^{3D})$, which are in subexponential time and superpolynomial time. According to Ladner's result [27], there exist NPI problems, which were constructed by removing strings of certain lengths from NP-complete languages. In the spin-glass case, it is understood that the NPI problems can be constructed by removing some interactions and/or spins from the AMC model. The computational complexity of the problems in the NPI region can be dealt with in quasi-polynomial times, for instance, O($N^{lgN}$), O($N^{lglgN}$), etc [61]. □

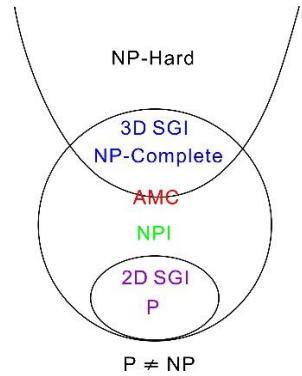

Figure 7 Phase diagram for the spin-glass Ising (SGI) model [61]. In the phase diagram, 3D SGI represents NP-complete problems, and P represents polynomial problems (2D SGI). NPI exists between NP-complete and P problems, while AMC is located on the border of NP-complete and NPI regions.

From the analysis above, we proposed a strategy for developing an optimum algorithm for calculations of physical properties (such as, the ground state, the free energy, the critical point, the phase transitions and the critical phenomena, etc.) of the spin-glass 3D Ising model [61].

1) Fix z-layers (z = 1, 2, 3, …) of the AMC model as an element of the algorithms, while performing a parallel computation of $l/z$ layers of this element.
2) Compare the precision as well as the accuracy of the results obtained by the above procedures, and determine the optimum value of z.

In this way, one can design the optimum algorithm to find/reach the exact solution with the sufficient accuracy and within the high precision in the shortest time. It can be improved greatly from the present status of $O(1.3^N)$ [47] to $O((1 + \varepsilon)^N)$ with $\varepsilon \to 0$ and $\varepsilon \neq 1/N$ [56,60,61], the best case if one can succeeded in the optimum value z = 1.. Since the spin-glass 3D Ising model is catalogued to NP-complete set, the optimum algorithm can be employed to compute the properties of other NP-complete problems (for instance, TSP, K-satisfiability problem, Knapsack problem, neural networks, etc.).

**3. Computational complexity of the traveling salesman problems**

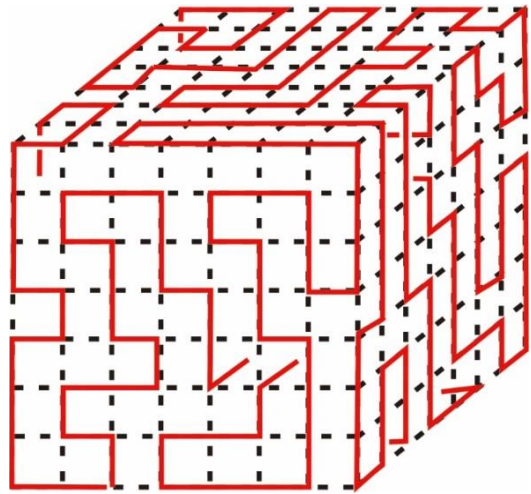

Figure 8. Schematic illustration of a TSP in a 3D lattice (with the lattice size $N = mnl$, here $m = n = l = 8$ as an example), $M_{TSP}^{3D}$. The black dashed lines represent the lattice, while the red solid lines represent the tour. There exist many possible crossings in the tour connecting all the lattice points (cities) in the 3D lattice. Since it is hard to illustrate the details of all the connections in the whole 3D lattice, we just illustrate the connections on the three planes outside.

**Theorem 7.** *The TSP in a 3D lattice, $M_{TSP}^{3D}$, can be mapped into a spin-glass 3D Ising model, $M_{SGI}^{3D}$, which is NP-complete. Namely, $M_{TSP}^{3D} \Leftrightarrow M_{SGI}^{3D}$ and $C(M_{TSP}^{3D}) = C(M_{SGI}^{3D})$.*

**Proof.** The (generalized) TSP can be defined as the following way

[2,5,10,12,14,15,20,32,37,43]: Suppose there are N points (cities) located at points $R_i$, i = 1, 2 ..... N, in a d-dimensional space. The nodes of a graph G = (N, E) represent N cities that must be visited by a salesman and the edges E represent roads or other transportation links connecting the cities. So, this problem can be mapped also to a restricted maximum cut (RMC) with a graph G = (V, E) with V is the set of vertices and E the set of undirected edges [1,51]. From and to one of the cities, the traveling salesman starts his tour and must return. The TSP asks for the minimum of $l(H) = \sum_{e \in H} l(e)$ for finding a Hamilton cycle (N,H) of G, where the function $l(e)$ associates the length to each edge $e \in E \rightarrow \mathbb{R}$. The tours are labeled by $t$. Let $L_t$ be the length of a tour $t$. A partition function is defined for TSP:

$$Z = \sum_t exp(-\beta L_t)$$

(11)

where $\beta = 1/(k_B T)$ is a parameter to be called the inverse temperature.

In order to show the mapping, we introduce auxiliary spin variables $S_i$ associated either with the cities or with the links connecting two cities. The various possible spin configurations generate various allowed tours. The length of a tour $t$, $L_t \equiv E\{S_i\}$, is a function of the spin variables, which is called the cost function. The cost function can be written as [1,51]:

$$E\{S_i\} = \sum_{(v_1,v_2) \in E} S_{v_1} S_{v_2}$$

(12)

where $S_i$ are Ising spins. If we assign an instance (i.e., configuration of coupling $J_{ij}$) of the spin-glass 3D Ising model to every instance (i.e., graph G = (N, E)) in TSP (or graph G = (V, E) in RMC), minimizing the cost function of the TSP (or RMC) is equivalent to minimizing the energy function of the spin-glass 3D Ising model [1,51]. This proves that solving TSP (or RMC) in a 3D space is at least as hard as finding the ground state of the spin-glass 3D Ising model, and hence belongs to the class of NP-complete problem [1,51]. The partition function of the spin system is [1,5,33,36,45,50]:

$$Z = \sum_t exp(-\beta L_t) = Trexp(-\beta E\{S_i\})$$

(13)

If one obtains a set of variables $S_i$ achieved above, one will realize a representation. Two representations can be provided for the TSP, in terms of (1) continuous field variables defined on cities, and (2) permutation group elements [5,33,36,45,50]. The question raised for the TSP can be answered in principle by calculating spin correlation functions and a proper understanding of the spin problem. The computational complexity of the TSP relies on its dimensionally. If the edges E of the TSP are limited in a 2D graph (without the crossings of the roads), the problem will be a P-problem; If the edges E of the TSP have the crossings with a 3D graph, the problem will be NP-complete.

The computational complexity of the spin-glass Ising models also relies on its dimensionally. In the previous work [3,4], the spin-glass 2D Ising model was proven to be a P-problem, whereas the spin-glass 3D Ising model was proven to be a NP-complete problem. Finding the ground state energy for an arbitrary set of couplings $J_{ij}$ in a spin-glass 3D Ising model on a cubic graph (and even a graph with a two-level grid [3,19,51]) is a hard combinatorial optimization task which in this case belongs to the class of NP-complete problems [1,3,18,34,40,47]. For more details, the readers refer to some reports on the relation between the spin-glass Ising models and the TSP [5,33,36,45,50]. Therefore, the TSP in a 3D lattice can be mapped into a spin-glass 3D Ising model which is NP-complete. Namely, the TSP in a 3D lattice can be reduced to a spin-glass 3D Ising model in polynomial time, vice versa. We have $M_{TSP}^{3D} \Leftrightarrow M_{SGI}^{3D}$ and $C(M_{TSP}^{3D}) = C(M_{SGI}^{3D})$. □

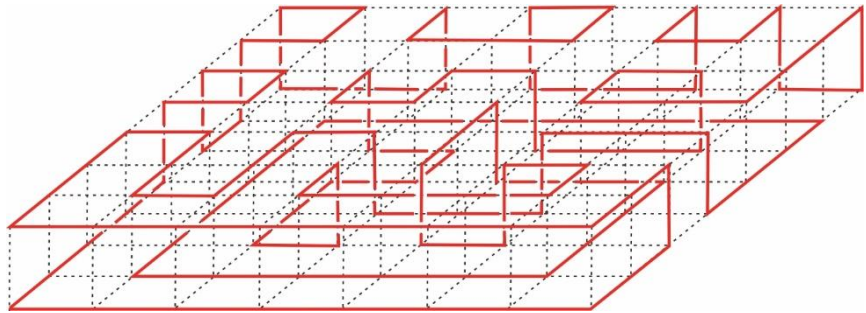

Figure 9. Schematic illustration of a TSP model on a two-level grid lattice (with the lattice size $N = mnl$, here $m = n = 8$ and $l = 2$), $M_{l=2,TSP}^{3D}$. The black dashed lines represent the lattice, while the red solid lines represent the tour. Here we are able to illustrate a tour as an example to connect all the lattice points (cities) in the two-layers ($l = 2$). There exist some crossings in the tour, which represent the character of the 3D space. In order to illustrate the connections, some solid lines are drawn to be not fitted

with the dashed lines for the two o-level grid lattice.

**Theorem 8.** $M_{l=2,TSP}^{3D} \Leftrightarrow M_{l=2,SGI}^{3D}$ is NP-complete.

**Proof.** Theorem 7 states that $M_{TSP}^{3D} \Leftrightarrow M_{SGI}^{3D}$. With the same mapping, we shall have $M_{l=2,TSP}^{3D} \Leftrightarrow M_{l=2,SGI}^{3D}$ According to Theorem 2, $M_{l=2,SGI}^{3D}$ is NP-complete. Thus, $M_{l=2,TSP}^{3D}$ is NP-complete. □

**Theorem 9.** *An AMC model, $M_{AMC,TSP}^{3D}$, exists in the TSP for the lower bound of its computational complexity, which is equivalent to the AMC model in the spin-glass 3D Ising model, $M_{AMC,SGI}^{3D}$. Namely, $M_{AMC,TSP}^{3D} \Leftrightarrow M_{AMC,SGI}^{3D}$ and $C(M_{AMC,TSP}^{3D}) = C(M_{AMC,SGI}^{3D})$.*

**Proof.** According to Theorem 7, $M_{TSP}^{3D} \Leftrightarrow M_{SGI}^{3D}$. According to Theorem 3, an AMC model exists in the spin-glass 3D Ising model. With the same mapping, we shall have $M_{AMC,TSP}^{3D} \Leftrightarrow M_{AMC,SGI}^{3D}$. Theorem 9 is validated as an immediate result of Theorems 3 and 7. □

**Theorem 10.** *The lower bound of the TSP in the 3D lattice is the computational complexity by brute force search of the AMC model for the TSP or the spin-glass 3D Ising model. Namely, $C_L(M_{TSP}^{3D}) = C_L(M_{SGI}^{3D}) \geq C^U(M_{AMC,SGI}^{3D}) = C^U(M_{AMC,TSP}^{3D})$.*

**Proof.** Theorem 7 proves that $C_L(M_{TSP}^{3D}) = C_L(M_{SGI}^{3D})$. Theorem 9 verifies that $C^U(M_{AMC,TSP}^{3D}) = C^U(M_{AMC,SGI}^{3D})$. According to Theorem 3 of [60], $C_L(M_{SGI}^{3D}) \geq C^U(M_{AMC,SGI}^{3D})$, thus we have $C_L(M_{TSP}^{3D}) \geq C^U(M_{AMC,TSP}^{3D})$. □

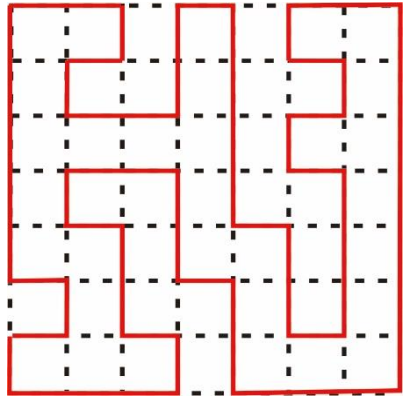

Figure 10. Schematic illustration of a TSP in a 2D lattice (with the lattice size $N = mnl$, here $m = n = 8$ and $l = 1$ as an example), $M_{TSP}^{2D}$. The black dashed lines represent the lattice, while the red solid lines represent the tour. There are no crossings in the tour connecting all the lattice points (cities) in the 2D lattice.

**Theorem 11.** *The TSP in a 2D lattice, $M_{TSP}^{2D}$, can be mapped into a spin-glass 2D Ising model, $M_{SGI}^{2D}$, which is P-problem [4]. Namely, $M_{TSP}^{2D} \Leftrightarrow M_{SGI}^{2D}$ and $C(M_{TSP}^{2D}) = C(M_{SGI}^{2D})$.*

**Proof.** It was proven in [4]. □

Figure 9 schematically illustrates a TSP model on a two-level grid lattice, $M_{l=2,TSP}^{3D}$, while Figure 10 illustrates a TSP model on a 2D lattice. In the last section, we have drawn Figures 2 and 3 to show the difference between the two-level grid spin-glass 3D Ising model, $M_{l=2,SGI}^{3D}$ and an AMC model for the spin-glass 3D Ising model, $M_{AMC,SGI}^{3D}$. The dashed line on the top layer ($l = 2$) in Figure 3 shows that it is a neighboring plane for the bottom layer with $l = 1$, to interact with along the third dimension, but without any intralayer interactions. It mean that one has to remove a spin-glass 2D Ising model, $M_{SGI}^{2D}$, from $M_{l=2,SGI}^{3D}$ to obtain $M_{AMC,SGI}^{3D}$. However, it is difficult to distinguish a TSP model on a two-level grid lattice, $M^{3D}_{l=2,TSP}$, and an AMC model for the TSP, $M^{3D}_{AMC,TSP}$, by schematic illustrations, since both of them need a two-level grid lattice for representation, in order to connect the lattice points (cities) by crossings. Nevertheless, by imaginary, we may process the same procedure to remove a 2D TSP model, $M^{2D}_{TSP}$, from $M^{3D}_{l=2,TSP}$ to obtain $M^{3D}_{AMC,TSP}$. This means that in $M^{3D}_{AMC,TSP}$, the intralayer tour connecting the lattice points (cities) on the top layer ($l=2$) of the two-level grid TSP lattice should be reduced to be as least as possible. It would be easier to distinguish the two cases by the formula in the following theorem.

**Theorem 12.** $M^{3D}_{AMC,TSP} = M^{3D}_{l=2,TSP} - M^{2D}_{TSP}$, *which is NP-complete.*

**Proof.** Theorem 8 verifies that $M^{3D}_{l=2,TSP} \Leftrightarrow M^{3D}_{l=2,SGI}$. Theorem 11 certifies $M^{2D}_{TSP} \Leftrightarrow M^{2D}_{SGI}$. Theorem 9 proves that $M^{3D}_{AMC,TSP} \Leftrightarrow M^{3D}_{AMC,SGI}$. According to Theorem 5, the identification, $M^{3D}_{AMC,SGI} = M^{3D}_{l=2,SGI} - M^{2D}_{SGI}$, is held, which is NP-complete. Therefore, the identification, $M^{3D}_{AMC,TSP} = M^{3D}_{l=2,TSP} - M^{2D}_{TSP}$, is valid, which is NP-complete. □

**Theorem 13.** $M^{3D}_{AMC,SGI}$ *is between* $M^{3D}_{SGI}$ *and* $M^{2D}_{SGI}$, *while* $M^{3D}_{AMC,TSP}$ *is between* $M^{3D}_{TSP}$ *and* $M^{2D}_{TSP}$.

**Proof.** This Theorem is validated evidently, following the results in Theorems 1, 4, 5, 7, 11, and 12. □

**Theorem 14.** *The lower bound of the computational complexity of the TSP in the 3D lattice,* $C_L(M^{3D}_{TSP})$ *or* $C^U(M^{3D}_{AMC,TSP})$, *is in subexponential and superpolynomial.*

**Proof.** According to Theorem 3 in [54], the computational complexity of the AMC model of a spin-glass 3D Ising model, $C(M^{3D}_{AMC})$, cannot be reduced to be less than

$O(2^{mn})$ by any algorithms. The AMC model must be computed by brute force search in order to obtain the exact solution of the spin-glass 3D Ising model. $C^U(M_{AMC,SGI}^{3D}) = O(2^{mn})$, and as $N \to \infty$, $O(2^{mn}) \to O((1+\varepsilon)^N)$ with $\varepsilon \to 0$ and $\varepsilon \neq 1/N$, which is subexponential, and superpolynomial. Theorem 10 states that $C_L(M_{TSP}^{3D}) = C_L(M_{SGI}^{3D}) \geq C^U(M_{AMC,SGI}^{3D}) = C^U(M_{AMC,TSP}^{3D})$. Thus, $C_L(M_{TSP}^{3D})$ or $C^U(M_{AMC,TSP}^{3D})$ is in subexponential and superpolynomial. □

**Theorem 15.** $M_{AMC,TSP}^{3D}$ is the border between $M_{TSP}^{3D}$ and $M_{NPI,TSP}$.

**Proof.** According to Theorem 6, there must exist a NPI problem $M_{NPI,SGI}$ for spin-glass Ising models, which is in between $M_{SGI}^{3D}$ and $M_{SGI}^{2D}$, and thus $M_{AMC,SGI}^{3D}$ is the border between $M_{SGI}^{3D}$ and $M_{NPI,SGI}$. Similarly, there must exist a NPI problem $M_{NPI,TSP}$ for TSP, which is located in between $M_{TSP}^{3D}$ and $M_{TSP}^{2D}$. Thus $M_{AMC,TSP}^{3D}$ is the border between $M_{SGI}^{3D}$ and $M_{NPI,TSP}$. □

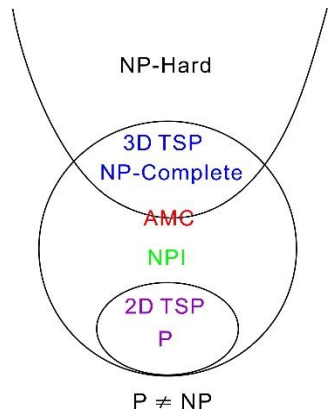

Figure 11 Phase diagram for the TSP [61]. In the phase diagram, 3D TSP represents the NP-complete problems, and P represents polynomial problems (2D TSP). NPI exists between NP-complete and P problems, while AMC is located on the border of NP-complete and NPI regions.

Summarizing the results in [56,60,61] and the present work, we reach the main theorem:

**Main Theorem: NP ≠ P.**

## 4. Conclusion

In conclusion, we have proven that in the spin-glass 3D Ising model and the TSP in the 3D lattice, the nontrivial topological structures, the non-planarity graphs, the nonlocalit**ies** and/or the long-range spin entanglements exist, while randomness presents together with frustrations. The computational complexity of the TSP depends on the dimensionality of the space where N cities are located at. If all of the cities are located in a 2D space, and/or if the tours are limited in the 2D plane without any crossings, the TSP will be a P-problem. But if the cities are located in a 3D space, and/or if the tours are distributed in the 3D lattice with crossings, the TSP will be NP-complete. The AMC model exists in both the spin-glass and the TSP models for determining the lower bound of their computational complexity. Any algorithms, which use any approximations and/or break the nonlocalities and the long-range spin entanglements in the AMC models, cannot result in the exact solution of the spin-glass 3D Ising model and the 3D TSP model. The AMC models in both models are proven to be NP-complete, but they are located at the border between the NP-complete problems and the NPI problems. The lower bound of the computational complexity of the spin-glass 3D Ising model $C_L(M_{SGI}^{3D})$ is the computational complexity by brute force search of the AMC model $C^U(M_{AMC,SGI}^{3D})$, while the lower bound of the computational complexity of the TSP in a 3D lattice $C_L(M_{TSP}^{3D})$ is the computational complexity by brute force search of the AMC model for the TSP $C^U(M_{AMC,TSP}^{3D})$. Namely, $C_L(M_{TSP}^{3D}) = C_L(M_{SGI}^{3D}) \geq C^U(M_{AMC,SGI}^{3D}) = C^U(M_{AMC,TSP}^{3D})$. All of them are in subexponential and superpolynomial. This means that one cannot develop a polynomial algorithm for either the 3D spin glass problem or the TSP. It is expected that one would develop a "best" algorithm for either the 3D spin glass problem or the 3D TSP if one starts from looking into the AMC model (or the 2D bilayer case). That is to develop a subexponential algorithm for the NP-complete problems (such as, the 3D spin glass problem, TSP, SAT, etc.), for instance, by parallel calculations of many AMC models (or 2D bilayers). The present work provides some implications on numerical algorithms for the NP-complete problems and bridges the gap between mathematics, computer science, and physics, which provides better understanding and efficiency of solutions of related hard problems.

**Acknowledgements**

This work has been supported by the National Natural Science Foundation of China under grant numbers 52031014.

**Declaration of competing interest:** The author declares no competing interests.

**References:**

[1] C.P. Bachas, Computer-intractability of the frustration model of a spin glass, J. Phys. A: Math. Gen. 17 (1984) L709-L712.

[2] W. Banzhaf, The "molecular" traveling salesman, Biol. Cybern. 64 (1990) 7-14.

[3] F. Barahona, On the computational complexity of Ising spin glass models, J. Phys. A 15 (1982) 3241-3253.

[4] F. Barahona, R. Maynard, R. Rammal, J. P. Uhry, Morphology of ground states of two-dimensional frustration model, J. Phys. A 15 (1982) 673-699.

[5] G. Baskaran, Y.T. Fu, P.W. Anderson, On the statistical mechanics of the traveling salesman problem, J. Statistical Physics, 45 (1986) 1-25.

[6] B. Berger, T. Leighton, Protein folding in the hydrophobic-hydrophilic (HP) model is NP-complete, J. Computational Biology, 5 (1998) 27-40.

[7] K. Binder, A.P. Young, Spin glasses: Experimental facts, theoretical concepts, and open questions, Rev. Mod. Phys. 58 (1986) 801-976.

[8] G. Biroli, R. Monasson, M. Weigt, A variational description of the ground state structure in random satisfiability problems，Eur. Phys. J. B 14 (2000) 551-568.

[9] V.D. Blondel, J. N. Tsitsiklis, A survey of computational complexity results in systems and control, Automatica 36 (2000) 1249-1274.

[10] V. Černý, Thermodynamical approach to the traveling salesman problem: An efficient simulation algorithm, J. Optimization Theory Applications, 45 (I985) 41-51.

[11] S. Cook, in Proceedings of the 3rd Annual ACM Symposium on Theory of Computing (ACM, New York, 1971), p. 151.

[12] G. Cornuéjols, J. Fonlupt, D. Naddef, The traveling salesman problem on a graph and some related integer polyhedral, Mathematical Programming 33 (1985) 1-27.

[13] S.F. Edwards, P.W. Anderson, Theory of spin glasses, J. Phys. F: Metal Phys. 5 (1975) 965-974.

[14] F. Favata, R. Walker, A study of the application of Kohonen-type neural networks to the Travelling Salesman Problem, Biol. Cybern. 64 (1991) 463-468.

[15] J. C. Fort, Solving a combinatorial problem via self-organizing process: An application of the Kohonen algorithm to the traveling salesman problem, Biol. Cybern. 59 (1988) 33M0.

[16] S. Fortune, J. Hopcroft, J. Wyllie, The directed subgraph homeomorphism problem, Theor. Computer Sci. 10 (1980) I1l-121.

[17] S. Franz, M. Leone, Replica bounds for optimization problems and diluted spin systems, J. Stat. Phys., 111 (2003) 535-564.

[18] M. Garey, D. S. Johnson, Computers and Intractability; A guide to the Theory of NP-completeness (Freeman, San Francisco,1979).

[19] F. Green, More about NP-completeness in the frustration model of spin-glasses, OR Spektrum 9 (1987) 161-165.

[20] M. Grötschel, M.W. Padberg, On the symmetric traveling salesman problem I: Inequalities, Mathematical Programming 16 (1979) 265-280.

[21] E. Ising, Beitrag zur Theorie des Ferromagnetismus, Z Phys, 31 (1925) 253-258.

[22] S. Istrail, Universality of Intractability for the Partition Function of the Ising Model Across Non-Planar Lattices, Proceedings of the 32nd ACM Symposium on the Theory of Computing (STOC00), ACM Press, p. 87-96, Portland, Oregon, May 21-23, 2000.

[23] B. Kaufman, Crystal Statistics II: Partition function evaluated by spinor analysis, Phys. Rev. 76 (1949) 1232-1243.

[24] J.B. Kogut, An introduction to lattice gauge theory and spin systems, Rev. Mod. Phys. 51 (1979) 659-713.

[25] M.W. Krentel, The complexity of optimization problems, J. Computer and System Sci. 36 (1988) 490-509.

[26] M.K. Kwan, Graphic programming using odd or even points, Chinese Mathematics, 1 (1962) 273-277.

[27] R. Ladner, On the structure of polynomial time reducibility, J. ACM 22 (1975) 155–171.

[28] L. Levin, Universal search problems (Russian: Универсальные задачи перебора, Universal'nye perebornye zadachi). Problems of Information Transmission (Russian: Проблемы передачи информации, Problemy Peredachi Informatsii). 9 (1973) 115-116.

[29] B.C. Li and W. Wang, Influence of a new long-range interaction on the magnetic properties of a 2D Ising layered model by using Monte Carlo method, Chinese Journal of Physics, 87 (2024) 525-539.

[30] B.C. Li and W. Wang, Exploration of dynamic phase transition of 3D Ising model with a new long-range interaction by using the Monte Carlo Method, Chinese Journal of Physics, Chinese Journal of Physics, **90** (2024) 15-30.

[31] S.L. Lou, S.H. Wu, Three-dimensional Ising model and transfer matrices, Chinese J. Phys., 38 (2000) 841-854.

[32] M. Mahia, Ö.K. Baykanb, H. Kodazb, A new hybrid method based on particle swarm optimization, ant colony optimization and 3-opt algorithms for traveling

salesman problem, Applied Soft Computing, 30 (2015) 484-490.

[33] O.C. Martin, R. Monasson, R. Zecchina, Statistical mechanics methods and phase transitions in optimization problems, Theoretical Computer Science 265 (2001) 3–67.

[34] M. Mézard, G. Parisi, M.A. Virasoro, Spin Glass Theory and Beyond (World Scientific, Singapore, 1987).

[35] M. Mézard, R. Zecchina, Random K-satisfiability problem: From an analytic solution to an efficient algorithm, Phys. Rev. E 66 (2002) 056126.

[36] R. Monasson, R. Zecchina, Statistical mechanics of the random K-satisfiability model, Phys. Rev. E 56 (1997) 1357-1370.

[37] C.C. Murray, A.G. Chu, The flying sidekick traveling salesman problem: Optimization of drone-assisted parcel delivery, Transportation Res. C 54 (2015) 86-109.

[38] L. Onsager, Crystal statistics I: A two-dimensional model with an order-disorder transition, Phys. Rev. 65 (1944) 117-149.

[39] C.H. Papadimitriou, The Euclidean traveling salesman problem is NP-complete, Theor. Computer Sci. 4 (1977) 237-244.

[40] C.H. Papadimitriou, K. Steiglitz, Combinatorial Optimization: Algorithms and Complexity, (Prentice-Hall, EnglewoodCliFs, NJ, 1982).

[41] C.H. Papadimitriou, Computational Complexity (Addison-Wesley, Reading, MA, 1994).

[42] J.H.H. Perk, Comment on 'Conjectures on exact solution of three-dimensional

(3D) simple orthorhombic Ising lattices', Phil. Mag. 89 (2009) 761-764.

[43] J.A.M. Potters, I.J. Curiel, S.H. Tijs, Traveling salesman games, Mathematical Programming 53 (1992) 199-211.

[44] F. Ricci-Tersenghi, M. Weigt, R. Zecchina1, Simplest random K-satisfiability problem, Phys. Rev. E 63 (2001) 026702.

[45] G.E. Santoro and E. Tosatti, Optimization using quantum mechanics: quantum annealing through adiabatic evolution, J. Phys. A: Math. Gen. 39 (2006) R393–R431.

[46] R. Savit, Duality in field theory and statistical systems, Rev. Mod. Phys. 52 (1980) 453-487.

[47] D.L. Stein, C.M. Newman, Spin Glasses and Complexity, (Princeton University Press, Princeton & Oxford, 2010).

[48] O. Suzuki, Z.D. Zhang, A method of Riemann-Hilbert problem for Zhang's conjecture 1 in a ferromagnetic 3D Ising model: trivialization of topological structure, Mathematics, 9 (2021) 776.

[49] C.A. Tovey, A simplified NP-complete satisfiability problem, Discrete Appl. Math. 8 (1984) 85-89.

[50] G. Venkataraman, G. Athithan, Spin glass, the travelling salesman problem, neural networks and all that, Pramāna - J. Phys., 36 (1991) 1-77.

[51] M. Yannakakis, Node-and-Edge Deletion, NP-complete Problems in Proc. 10th Annual ACM Symp. on Theory of Computing (New York: Association for Computing Machinery), (1978) p 253

[52] Z.D. Zhang, Conjectures on the exact solution of three - dimensional (3D) simple

orthorhombic Ising lattices, Phil. Mag. 87 (2007) 5309-5419.

[53] Z.D. Zhang, Mathematical structure of the three - dimensional (3D) Ising model, Chinese Phys. B 22 (2013) 030513.

[54] Z.D. Zhang, The nature of three dimensions: Non-local behavior in the three-dimensional (3D) Ising model, J. Phys. Conf. Series 827 (2017) 012001.

[55] Z.D. Zhang, O. Suzuki, N.H. March, Clifford algebra approach of 3D Ising model, Advances in Applied Clifford Algebras, 29 (2019) 12.

[56] Z.D. Zhang, Computational complexity of spin-glass three-dimensional (3D) Ising model, J. Mater. Sci. Tech. 44 (2020) 116-120.

[57] Z.D. Zhang, Exact solution of two-dimensional (2D) Ising model with a transverse field: a low-dimensional quantum spin system, Physica E 128 (2021) 114632.

[58] Z.D. Zhang, O. Suzuki, A method of the Riemann-Hilbert problem for Zhang's conjecture 2 in a ferromagnetic 3D Ising model: topological phases, Mathematics, 9 (2021) 2936.

[59] Z.D. Zhang, Topological quantum statistical mechanics and topological quantum field theories, Symmetry, 13 (2022) 323.

[60] Z.D. Zhang, Mapping between spin-glass three-dimensional (3D) Ising model and Boolean satisfiability problems, Mathematics, 11 (2023) 237.

[61] Z.D. Zhang, Lower bound of computational complexity of Knapsack problems, AIMS Math. 10 (2025) 11918-11938.